\documentclass[reprint,onecolumn,secnumarabic,amssymb, nobibnotes, aps, prfluids]{revtex4-2}

\usepackage{psfrag}

\usepackage{xcolor}
\usepackage{bbm}

\usepackage{lmodern}
\usepackage[T1]{fontenc}
\usepackage[utf8]{inputenc}
\usepackage{indentfirst}
\usepackage{color}
\usepackage{microtype}
\usepackage{lipsum}	
\usepackage{parskip}
\usepackage{setspace}

\usepackage{graphicx}
\usepackage{subcaption}
\usepackage{epstopdf}
\usepackage{epsfig}
\usepackage{rotating}
\usepackage{morefloats}

\usepackage{amsmath}
\usepackage{amssymb}
\usepackage{bm}
\usepackage{mnsymbol}
\usepackage{wasysym}
\usepackage{icomma}
\usepackage{multirow}
\usepackage{placeins}
\usepackage{mathrsfs}
\usepackage{mathtools}
\usepackage{subeqnarray}

\usepackage[english]{babel}
\usepackage{bibentry}
\usepackage{hyperref}
\usepackage{cleveref}
\usepackage{natbib}

\usepackage{etoolbox}
\makeatletter
\patchcmd\affils@present@group
 {\frontmatter@affiliationfont}
 {\frontmatter@affiliationfont\begingroup\lccode`~=`,\lowercase{\endgroup\let~}\active@comma}
 {}{}
\makeatother

\usepackage{makecell}

\begin{document}


\title{LES-informed resolvent-based estimation of turbulent pipe flow}


\author{Filipe R. do Amaral}
\thanks{filipe.ramos.do.amaral@univ-poitiers.fr}
\affiliation{D\'{e}partement~Fluides,~Thermique,~Combustion,~Institut~Pprime-CNRS-Universit\'{e}~de~Poitiers-ENSMA,~86000~Poitiers,~France}
\author{Andr\'{e} V. G. Cavalieri}
\thanks{andre@ita.br}
\affiliation{Divis\~{a}o~de~Engenharia~Aeroespacial,~Instituto~Tecnol\'{o}gico~de~Aeron\'{a}utica,~12228-900~S\~{a}o~Jos\'{e}~dos~Campos,~SP,~Brazil}
\date{\today}
\maketitle


A resolvent-based methodology is employed to obtain spatio--temporal estimates of turbulent pipe flow from probe measurements of wall shear-stress fluctuations.
Direct numerical simulations (DNS) and large-eddy simulations (LES) of turbulent pipe flow  at friction Reynolds number of 550 are used as databases.
We consider a DNS database as the true spatio--temporal flow field, from which wall shear-stress fluctuations are extracted  and considered as measurements.
A resolvent-based estimator is built following our earlier work (Amaral et al. \emph{J. Fluid Mech.}, vol. 927, 2021, p. A17), requiring a model for the nonlinear (or forcing) terms of the Navier-Stokes equations system, which are obtained from another DNS database, as in our earlier work, and from a series of computationally cheaper LES databases with coarser grids; the underlying idea is that LES may provide accurate statistics of non-linear terms related to large-scale structures at a low computational cost.
Comparisons between the DNS and the estimates indicate that sufficiently accurate results can be achieved with estimators built with statistics from LES with an order of magnitude less grid points than the DNS, with estimates closely matching the reference DNS results up to the buffer layer and reasonable agreement up to the beginning of the log layer.

\section{Introduction}
\label{sec:introduction}

Estimation of space--time flow fluctuations from noisy, low-rank measurements is an interesting option for the understanding of the turbulence physics, design of flow control strategies and reconstruction of missing or corrupted data.
For wall-bounded turbulent flows, wall quantities such as shear stress and/or pressure are usually employed as inputs for the estimation algorithms, as for practical applications the measurement of such quantities is easier to obtain than, e.g. the velocity components at a given distance from the wall.
To build the estimator, model-based methodologies can be used \citep{bewley2004skin, hoepffner2005state, chevalier2006state, colburn2011state, illingworth2018estimating, wang2022what}, although it is also possible to conduct flow estimations based solely on data \citep{encinar2019logarithmic, sasaki2019transfer, guastoni2021convolutional, nekkanti2023gappy}.
For both model- and data-driven methodologies the basic idea is to obtain transfer functions between the measurements and the estimated flow state, bearing in mind that the model-based methodologies have the additional advantage of providing insight on the underpinning physics.

In recent years, the use of linear models to understand the physics that drive turbulent wall-bounded flows has become widespread.
Linear models provide a simple framework to work with and the emergence of tools such as resolvent analysis enables modeling coherent structures and self-sustaining mechanisms in flows \citep{mckeon2017engine, oehler2018linear, tissot2021stochastic, herrmann2021data, lopezdoriga2022resolvent, symon2022eddy}.
In the resolvent framework the Navier-Stokes system is written in the state--space form, and the nonlinear terms are interpreted as external forcing terms \citep{mckeon2010critical, hwang2010linear, beneddine2016conditions, taira2017modal}, hence providing a convenient input--output formulation, relating the flow response and the forcing modes, related to non-linear terms in the Navier-Stokes system.

The input--output formulation enables the use of control theory tools \citep{bagheri2009input}, that can be adapted to estimate the flow state components after low-rank measurements.
\citet{towne2020resolvent} introduced a resolvent-based estimator for flow statistics, which was further generalized by \citet{martini2020resolvent} for time-domain estimates.
For the latter case, in order to build the transfer functions between the low-rank measurements and the flow state components, it is necessary inform the algorithm with the cross-spectral density (CSD) of the nonlinear terms of the Navier-Stokes system, treated as forcing.
If the true forcing CSD is used, optimal estimates of time-varying flow quantities are obtained.
Other forcing models provide sub-optimal estimates.
Such estimates are not causal, as the full time series of sensor data is required for estimation; extension to causal estimation, using only past sensor information, is proposed by \citep{martini2022resolvent}.
In our previous work we have successfully applied the methodology developed by \citet{martini2020resolvent} to direct numerical simulation (DNS) of turbulent channel flow, using wall shear-stress and pressure low-rank measurements \citep{amaral2021resolvent}.
Results show a close agreement between estimates and reference DNS fluctuations in the near wall region, and good agreement for large scale structures throughout the channel.
A key feature is the use of forcing statistics extracted from the DNS database, which leads to an optimal estimator but requires expensive simulation and post-processing to obtain the forcing CSD.
One option to model the forcing statistics is to consider it as spatially white noise.
\citet{amaral2021resolvent} show that this is a good choice for near-wall structures estimates, close to where the measurements were taken, whereas far from the wall, the estimator failed to deliver good results.
The work also explored the use of a standard Cess eddy-viscosity model \citet{cess1958survey, delalamo2006linear} embedded within the linear operator to somehow account for the nonlinear terms that are missing in the linearized Navier-Stokes (LNS) system.
The eddy-viscosity improved the channel flow large-scale structures estimates but at the cost of worsening the near-wall structures estimates.

\citet{chinta2022statiscally} employed the resolvent framework to estimate the velocity field of turbulent channel flow after low-rank measurements.
Their method differs from that by \citet{martini2020resolvent} as they compute the resolvent modes for various wavenumber/frequency combinations and then assume that the flow state is a linear combination of such modes, calibrating linear coefficients after the input/measurements data.
In other words, the method by \citet{chinta2022statiscally} seeks to identify the resolvent modes that best represent the measurement data.
It is interesting to note that the inclusion of an eddy-viscosity model in the linear operator also had a dual effect in the results by \citet{chinta2022statiscally}: although it improved the flow statistics, it also increased the velocity field estimates errors.

Obtaining optimal resolvent-based estimators to turbulent flow was shown to be feasible in \citet{amaral2021resolvent}, but with a high computational cost related to the extraction of forcing statistics from a DNS database.
In the present paper we employ the \citet{martini2020resolvent} resolvent-based methodology to estimate the space--time velocity fluctuation components of turbulent pipe flow at friction Reynolds number $Re_{\tau} \approx 550$ using wall shear-stress measurements.
In addition to DNS, we also explore the capability of wall-resolved large-eddy simulations (LES) to construct estimators.
A first DNS database is used as the reference case from which we extract low-rank measurements of wall-shear stresses.
A second DNS and the other LES databases provide the forcing (nonlinear) statistics to build the linear estimators.
The underlying assumption is that LES provides sufficiently accurate statistics of large-scale structures, including associated non-linear terms, and thus may be used to construct estimators with near-optimal performance.
Here we investigate the capability of such LES databases on the reconstruction of the space--time flow field, aiming to obtain a reliable and lower-cost estimator that could be used for various high-Reynolds-number flows of practical interest.

The remainder of the manuscript is organized as follows.
Section \ref{sec:methodology} presents the methods employed in this work, including the resolvent-based estimator algorithm and details on the turbulent pipe flow simulations.
Section \ref{sec:results} contains the results, including a direct comparison among the reference DNS and the estimators using the different strategies to model the nonlinear terms of the LNS equations, as well as metrics to evaluate the performance of each estimator.
Finally, section \ref{sec:conclusions} presents the conclusions of this study.

\section{Methodology}
\label{sec:methodology}

\subsection{Resolvent-based estimator}

Let us begin by applying a Reynolds decomposition over the flow state, i.e. $\boldsymbol{q} = \boldsymbol{\bar{q}} + \boldsymbol{q^{\prime}}$, where $\boldsymbol{\bar{q}}$ and $\boldsymbol{q^{\prime}}$ denote mean flow and fluctuation around the mean flow components.
In the resolvent framework, a the forcing term is obtained gathering all nonlinear terms on $\boldsymbol{q}^{\prime}$.
The vector of velocity components is given by $\boldsymbol{u} = \left[u_x~u_r~u_\theta\right]$, with $u_x$, $u_r$ and $u_\theta$ as the streamwise, radial and azimuthal velocity components, respectively.
The full state vector is written as
\begin{equation}
	\boldsymbol{q} = \left[\boldsymbol{u}~p\right]^{T} \mbox{,}
	\label{eq:state_components}
\end{equation}
\noindent with $\boldsymbol{q} = \boldsymbol{q}\left(x,r,\theta,t\right)$, where $r$ and $\theta$ indicate the radial and azimuthal directions, respectively, $t$ denotes time and $p$ the pressure component.

The linear Navier-Stokes (LNS) equations in cylindrical coordinates can be written as
\begin{subeqnarray}
	{\partial_t \boldsymbol{u}} + {u_r \partial_r \bar{U} \boldsymbol{e_x}} + {\bar{U} \partial_x \boldsymbol{u}} &=& {\boldsymbol{\nabla} p} + \frac{1}{Re} {\boldsymbol{\nabla}^2 \boldsymbol{u}} + \boldsymbol{f} \mbox{,} \\
	\boldsymbol{\nabla} \cdot \boldsymbol{u} &=& 0 \mbox{,}
	\label{eq:LNS}
\end{subeqnarray}
\noindent where $\partial$ denotes partial derivatives with respect to $t$, $r$ or $x$ for time, radial and streamwise directions, respectively, $\bar{U}$ is the mean turbulent velocity profile, $\boldsymbol{\nabla}$ is the gradient operator, $Re$ is the Reynolds number based on bulk velocity, $\boldsymbol{f}$ denotes the forcing components and $\boldsymbol{e_x}$ is the unity vector in the streamwise direction.
For use in resolvent analysis, the forcing $\boldsymbol{f}$ includes the non-linear terms in the Navier-Stokes equation, $\boldsymbol{f} = \left(-\boldsymbol{u} \cdot \nabla \boldsymbol{u} \right)$, such that equations \ref{eq:LNS} are an exact rearrangement of the full Navier-Stokes system.

Regarding the forcing components, they are structured as
\begin{equation}
	\boldsymbol{f} = \left[f_x~f_r~f_\theta\right]^{T} \mbox{,}
	\label{eq:forcing_components}
\end{equation}
\noindent with $\boldsymbol{f} = \boldsymbol{f}\left(x,r,\theta,t\right)$.
Only molecular viscosity is considered in the present work, such that Eq. \ref{eq:LNS} is exact if the full forcing $\boldsymbol{f}$ is used \citep{morra2021colour}.
In the following, primes ($^\prime$) will be dropped from the state and forcing notations for simplification.

In a discretized state-space form, considering a grid with $N_r$ points in the radial direction, the LNS equations take the form
\begin{subeqnarray}
	\boldsymbol{M} \frac{d \boldsymbol{q}(t)}{d t} &=& \boldsymbol{A} \boldsymbol{q}(t) + \boldsymbol{B} \boldsymbol{f} \mbox{,} \\
	\boldsymbol{z}(t) &=& \boldsymbol{C} \boldsymbol{q}(t) + \boldsymbol{n}(t) \mbox{.}
	\label{eq:state-space_time}
\end{subeqnarray}
\noindent where $\boldsymbol{A}$ denotes the linearized Navier-Stokes operator, $\boldsymbol{B}$ is the input matrix that restricts the forcing terms to appear only in the momentum equation, $\boldsymbol{z}$ is the system observation (measurements), $\boldsymbol{C}$ is the observation matrix that selects $N_{s}$ sensor readings from the state vector (in the present paper, wall-shear stresses in the axial and azimuthal directions), and $\boldsymbol{n}$ is the measurement noise.
$\boldsymbol{M}$ is a diagonal matrix whose entries are set to one and zero for the momentum and continuity equations, respectively.
The dependencies on wall-normal distance $y$ (or radial distance $r$), were dropped to simplify notations.

The state components can be written as a superposition of Fourier modes as
\begin{equation}
	\boldsymbol{q}\left(x,r,\theta\right) = \sum_{m} \sum_{\alpha} {\int_{-\infty}^{\infty} {\boldsymbol{\hat{q}}(\alpha,r,m,\omega) e^{i \left({\alpha x} + {m \theta} - {\omega t}\right)} d\omega}} \mbox{,}
	\label{eq:fourier}
\end{equation}
\noindent where $\omega$ denotes frequency, $\alpha$ and $m$ indicate longitudinal and azimuthal wavenumbers, respectively, hats are used for Fourier-transformed quantities, $m$ is constrained to be an integer number and $i = \sqrt{-1}$.
Similar to the azimuthal direction, a Fourier series is taken along $x$, since periodic boundary conditions are applied for the axial direction in the simulations considered here.

Equation \ref{eq:state-space_time} can be written in the frequency domain as
\begin{equation}
	\boldsymbol{\hat{z}}(\omega) = \left[\boldsymbol{C} (- i \omega \boldsymbol{M} - \boldsymbol{A})^{-1} \boldsymbol{B}\right] \boldsymbol{\hat{f}}(\omega) + \boldsymbol{\hat{n}}(\omega) \mbox{.}
	\label{eq:state-space_freq}
\end{equation}
\noindent where dependencies on longitudinal ($\alpha$) and azimuthal ($m$) wavenumbers were also dropped to simplify notations.

The term $\boldsymbol{R} = (- i \omega \boldsymbol{M} - \boldsymbol{A})^{-1} = \boldsymbol{L}^{-1}$ is the resolvent operator, which is well posed once non-slip boundary conditions are enforced for the three velocity components.
For pipe flow, the linear operator $\boldsymbol{A}$ is written in cylindrical coordinates \citep{luhar2014opposition} and linearization is around the mean turbulent profile, considered as known; the linearized operator becomes
\begin{equation}
	\boldsymbol{A} =
	\left[\begin{array}{cccc}
		- i \alpha \boldsymbol{\bar{U}} + \frac{\boldsymbol{\Delta} + \boldsymbol{r}^{-2}}{Re}	& - \boldsymbol{D} \boldsymbol{\bar{U}}	& \boldsymbol{Z}	& - i \alpha \boldsymbol{I} \\
		\boldsymbol{Z}	& - i \alpha \boldsymbol{\bar{U}} + \frac{\boldsymbol{\Delta}}{Re}	& -\frac{2 i m \boldsymbol{r}^{-2}}{Re}	& - \boldsymbol{D} \\
		\boldsymbol{Z}	& \frac{2 i m \boldsymbol{r}^{-2}}{Re} 	& -i \alpha \boldsymbol{\bar{U}} + \frac{\boldsymbol{\Delta}}{Re}	& - i m \boldsymbol{r}^{-1} \\
		i \alpha \boldsymbol{I}	& \boldsymbol{D} + \boldsymbol{r}^{-1}	& i m \boldsymbol{r}^{-1}	& \boldsymbol{Z}
	\end{array}\right] \mbox{,}
	\label{eq:linear_operator}
\end{equation}
\noindent where $\boldsymbol{\Delta} = - {\alpha^2 \boldsymbol{I}} - {\left(m^2 + 1\right) \boldsymbol{r}^{-2}} + {\boldsymbol{r}^{-1} \boldsymbol{D}} +\boldsymbol{D}^2$, $\boldsymbol{D} = \boldsymbol{\frac{d}{d r}}$ is a diagonal matrix with the radial direction derivative operator (finite differences scheme), $\boldsymbol{I}$ is the identity matrix, $\boldsymbol{r}$ and $\boldsymbol{\bar{U}}$ are diagonal matrices containing the radial discretization and mean turbulent velocity profile, respectively, and $\boldsymbol{Z}$ is the zero matrix.

The actuation/input operator $\boldsymbol{B}$ is defined as
\begin{equation}
	\boldsymbol{B} =
	\left[\begin{array}{ccc}
		\boldsymbol{I}	& \boldsymbol{Z}	& \boldsymbol{Z} \\
		\boldsymbol{Z}	& \boldsymbol{I}	& \boldsymbol{Z} \\
		\boldsymbol{Z}	& \boldsymbol{Z}	& \boldsymbol{I} \\
		\boldsymbol{Z}	& \boldsymbol{Z}	& \boldsymbol{Z}
	\end{array}\right] \mbox{,}
	\label{eq:actuation_matrix}
\end{equation}
\noindent whereas the observation/output operator $\boldsymbol{C}$ is given by
\begin{equation}
	\boldsymbol{C} =
	\left[\begin{array}{cccc}
		\boldsymbol{D}_{wall}	& \boldsymbol{Z}	& \boldsymbol{Z}				& \boldsymbol{Z} \\
		\boldsymbol{Z}				& \boldsymbol{Z}	& \boldsymbol{D}_{wall}	& \boldsymbol{Z}
	\end{array}\right] \mbox{,}
	\label{eq:observation_matrix}
\end{equation}
\noindent where $\boldsymbol{D}_{wall}$ means that only the derivatives correspondent to the pipe wall (i.e. the corresponding line of the differentiation matrix) are taken into account.
The diagonal matrix $\boldsymbol{M}$ is written as
\begin{equation}
	\boldsymbol{M} =
	\left[\begin{array}{cccc }
		\boldsymbol{I}	& \boldsymbol{Z}	& \boldsymbol{Z}	& \boldsymbol{Z} \\
		\boldsymbol{Z}	& \boldsymbol{I}	& \boldsymbol{Z}	& \boldsymbol{Z} \\
		\boldsymbol{Z}	& \boldsymbol{Z}	& \boldsymbol{I}	& \boldsymbol{Z} \\
		\boldsymbol{Z}	& \boldsymbol{Z}	& \boldsymbol{Z}	& \boldsymbol{Z}
	\end{array}\right] \mbox{,}
	\label{eq:m_matrix}
\end{equation}

From the definitions above, it is possible to obtain an optimal linear transfer function ($\boldsymbol{\hat{T}_q}$) between the system observation ($\boldsymbol{\hat{z}}$) and the estimated flow state components ($\boldsymbol{\hat{\tilde{q}}}$), such that
\begin{equation}
	\boldsymbol{\hat{\tilde{q}}} = \boldsymbol{\hat{T}_q} \boldsymbol{\hat{z}} \mbox{,}
	\label{eq:f_est}
\end{equation}
\noindent where $\boldsymbol{\hat{T}_q}$ is the transfer function and dependency on frequency $\omega$ was dropped to simplify notations.
The tilde superscript denotes estimates.

\citet{martini2020resolvent} derived a expression for $\boldsymbol{\hat{T}_q}$ that is based on the minimization of the error between the true ($\boldsymbol{\hat{f}}$) and estimated ($\boldsymbol{\hat{\tilde{f}}}$) forcing terms.
The resulting transfer function is given by
\begin{equation}
	\boldsymbol{\hat{T}_q} = \boldsymbol{R} \boldsymbol{B} \boldsymbol{P_{ff}} {\boldsymbol{H}}^{*} \left(\boldsymbol{H} \boldsymbol{P_{ff}} {\boldsymbol{H}}^{*} + \boldsymbol{P_{nn}}\right)^{-1} \mbox{,}
	\label{eq:Tq}
\end{equation}
\noindent where $\boldsymbol{H} = \boldsymbol{C} \boldsymbol{R} \boldsymbol{B}$ is the resolvent operator including the observation ($\boldsymbol{C}$) and actuation ($\boldsymbol{B}$) matrices, $\boldsymbol{P_{nn}} = \left\langle \boldsymbol{\hat{n}} \boldsymbol{\hat{n}}^{*} \right\rangle$ and $\boldsymbol{P_{ff}} = \left\langle \boldsymbol{\hat{f}} \boldsymbol{\hat{f}}^{*} \right\rangle$ are the CSDs of sensor noise and forcing, respectively.
The asterisk ($^*$) indicates a Hermitian transpose.

To build the transfer function, it is necessary to specify \emph{a priori} the forcing CSD ($\boldsymbol{P_{ff}}$).
When true forcing statistics are known, equation \ref{eq:Tq} provides the optimal linear estimator, whereas other models for the forcing CSD provide sub-optimal estimators.
In the present paper, we focus on the modeling of $\boldsymbol{P_{ff}}$ through DNS and LES, extracting the non-linear terms of the Navier--Stokes system direct from the simulations.
Many other strategies can be employed to address the forcing statistics in the context of linear modeling for turbulent flows as, e.g., entropy-based modeling \citep{zare2017colour} or resolvent-based proper orthogonal decomposition (RESPOD) \citep{karban2022self}, among others.
The reviews by \citet{zare2020stochastic} and \citet{jovanovic2021bypass} explore the importance of the forcing statistics, and means to obtain it, in the context of modeling and control of turbulent flows.

The measurements noise statistics ($\boldsymbol{P_{nn}}$) must also be considered to evaluate the transfer function, and non-zero noise in the sensor readings can reduce the estimation efficiency \citep{martini2020resolvent}.
In the cited work, it is shown that if the sensor readings are separated into a noiseless component and a noise component, the noiseless reading can only be recovered for small values of the measurement noise. 
Larger noise levels lead to smaller estimated components.
In the present study, as the data is obtained directly from the simulations, the measurement noise is virtually non-existent.
Nevertheless, such component regularizes the estimation, maintaining the problem well posed \citep{martini2020resolvent}.

The transfer function obtained through equation \ref{eq:Tq} is non-causal, although it is possible to extend this approach to causal estimations through the Wiener--Hopf formalism, as addressed by \citet{martini2022resolvent} and \citet{audiffred2023experimental}.
As the focus of this work is on the applicability of using LES forcing statistics to build estimators, optimal causal transfer functions will be left for future work.

The snapshots are reconstructed in space and time according to the procedures briefly addressed below.
First, it is necessary to take the inverse Fourier transform of the transfer function $\boldsymbol{\hat{T}_q}$, equation \ref{eq:Tq}, in order to return to time domain and obtain $\boldsymbol{T_q}$,
\begin{equation}
	\boldsymbol{T_q}(\alpha,r,m,t) = \int^{\infty}_{-\infty} \boldsymbol{\hat{T}_q}(\alpha,r,m,\omega) e^{i \omega t} d\omega \mbox{.}
	\label{eq:ifft_Tq_time}
\end{equation}
Hence, the time domain transfer function $\boldsymbol{T_q}$ must be convolved with the measurements/observations $\boldsymbol{z}$ to evaluate the state estimate in time domain $\boldsymbol{\tilde{q}}$,
\begin{equation}
	\boldsymbol{\tilde{q}}(\alpha,r,m,t) = \int^{\infty}_{-\infty} \boldsymbol{T_q}(\alpha,r,m,\tau) \boldsymbol{z}(\alpha,r,m,t-\tau) d\tau \mbox{.}
	\label{eq:q_time}
\end{equation}
Finally, double inverse Fourier transforms in the azimuthal and longitudinal directions are taken in order to return from wavenumber domain to physical space,
\begin{equation}
	\boldsymbol{\tilde{q}}(x,r,\theta,t) =\sum_{m} \sum_{\alpha} \boldsymbol{\tilde{q}}(\alpha,r,m,t) e^{i \alpha x + i m \theta} \mbox{.}
	\label{eq:ifft_ab}
\end{equation}
Note that although the notation $\tilde{q}$ is used in both sides of equation \ref{eq:ifft_ab}, in the left-hand side the estimated stated is function of the physical domain variables, i.e. $(x,r,\theta,t)$, whereas in the right-hand side, the state is function of the wavenumber domain variables, i.e. $(\alpha,r,m,t)$.

It is also possible to model the forcing statistics ($\boldsymbol{P_{ff}}$), which provides a cheaper and sub-optimal estimator. 
Another option is to consider an eddy-viscosity model in the linear operator $\boldsymbol{A}$, equation \ref{eq:linear_operator}, to somehow take into account the nonlinear terms of the Navier-Stokes system, as discussed by \citet{symon2021energy}, \citet{morra2021colour} and \citet{amaral2021resolvent}.
Appendix \ref{app:eddy_viscosity} addresses the linear operator containing an eddy-viscosity model.
In this case, the forcing statistics are considered as white-noise in space.
We will consider both cheap estimators to compare them with the LES ones.

\subsection{Numerical simulations}

\begin{figure}
	\centerline{\includegraphics[width=0.5\textwidth]{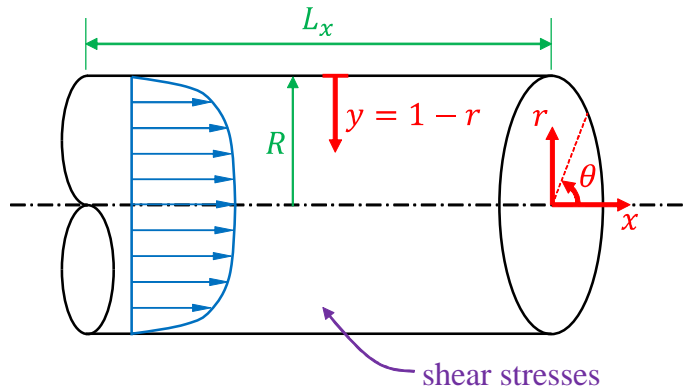}}
	\caption{Sketch of the channel flow geometry, dimensions (green), coordinate system (red), mean flow (blue) and wall measurements employed to perform the estimations (purple).}
	\label{fig:pipe_flow_sketch}
\end{figure}

To generate the databases, we conducted numerical simulations with the Openpipeflow code \citep{willis2017openpipeflow}.
Periodic boundary conditions were assumed in the streamwise and azimuthal directions.
Figure \ref{fig:pipe_flow_sketch} shows a sketch of the geometry and coordinate system employed in this study.

For all simulations the pipe length is $L_x = 10 R$, where $R$ is the pipe radius, and the bulk Reynolds number is $Re_b = \frac{U_b D}{\nu} = 19,000$, where $U_b$ is the bulk velocity, $D = 2 R$ is the pipe diameter and $\nu$ is the kinematic viscosity.
Table \ref{tab:simulations_parameters} shows the parameters for all cases, including the number of radial ($N_r$), streamwise ($N_x$) and azimuthal ($N_{\theta}$) grid points, the mesh discretization in the streamwise ($\Delta x^+$), azimuthal ($(R \Delta \theta)^+$) and radial ($\Delta r^+$) directions and the mesh points ratio with respect to the DNS case ($N/N_{DNS}$).
Plus symbols denote inner (wall and/or viscous) units.
The data extracted from the simulations to obtain the measurements, forcing components and comparisons with the reference DNS components contain 5,000 snapshots and the time steps based on outer units is $\Delta t = 0.1$.
Cases starting with D denote DNS, whereas letter L indicate LES, carried out using \citet{smagorinsky1963general} subgrid scale model, with a Smagorinsky constant set as $C_s = 0.05$; the wall damping function of van Driest \citep{vandriest1956turbulent} was used.
As estimations lose accuracy for large wavenumbers, only the lowest 16 and 32 streamwise and azimuthal wavenumbers were used to construct the estimators.
Welch's method \citep{welch1967fft} was employed to evaluate the forcing and state components statistics, with blocks containing $N_{fft} = 512$ time steps and 75\% overlap.
Appendix \ref{app:convergence} shows the results of the block size convergence test, justifying the use of $N_{fft} = 512$.
A Hann window was applied to each block to minimize spectral leakage.

\begin{table}
  \begin{center}
		\def~{\hphantom{0}}
		\caption{Numerical simulation parameters.}
		\label{tab:simulations_parameters}
		\begin{tabular}{c c c c c c c c c}
				Case	& $Re_{\tau}$	& $N_r$	& $N_x$	& $N_{\theta}$	& $\Delta x^+$	& $(R \Delta \theta)^+$	& $\Delta r^+$	& $N/N_{DNS}$	\\ [3pt]
				\hline
				D1		& 550.3				& 128 	& 528		& 528						& 10.4					& 6.5										& 0.07-6.3			& 1.000				\\
				D2		& 550.3				& 128 	& 528		& 528						& 10.4					& 6.5										& 0.07-6.3			& 1.000				\\
				L1		& 554.1				& 96 		& 288		& 288						& 18.9					& 11.9									& 0.4-14.7			& 0.223				\\
				L2		& 568.6				& 96 		& 192		& 192						& 28.6					& 18.0									& 0.4-15.2			& 0.099				\\
				L3		& 569.1				& 64 		& 96		& 96						& 59.3					& 37.2									& 0.7-22.2			& 0.017				\\
				L4		& 551.8				& 64 		& 64		& 64						& 86.1					& 54.1									& 0.7-22.9			& 0.007				\\
				L5		& 509.6				& 64 		& 32		& 32						& 159.3					& 100.1									& 0.7-22.9			& 0.002				\\
		\end{tabular}
  \end{center}
\end{table}

\begin{figure}
	\centerline{\includegraphics[width=\textwidth]{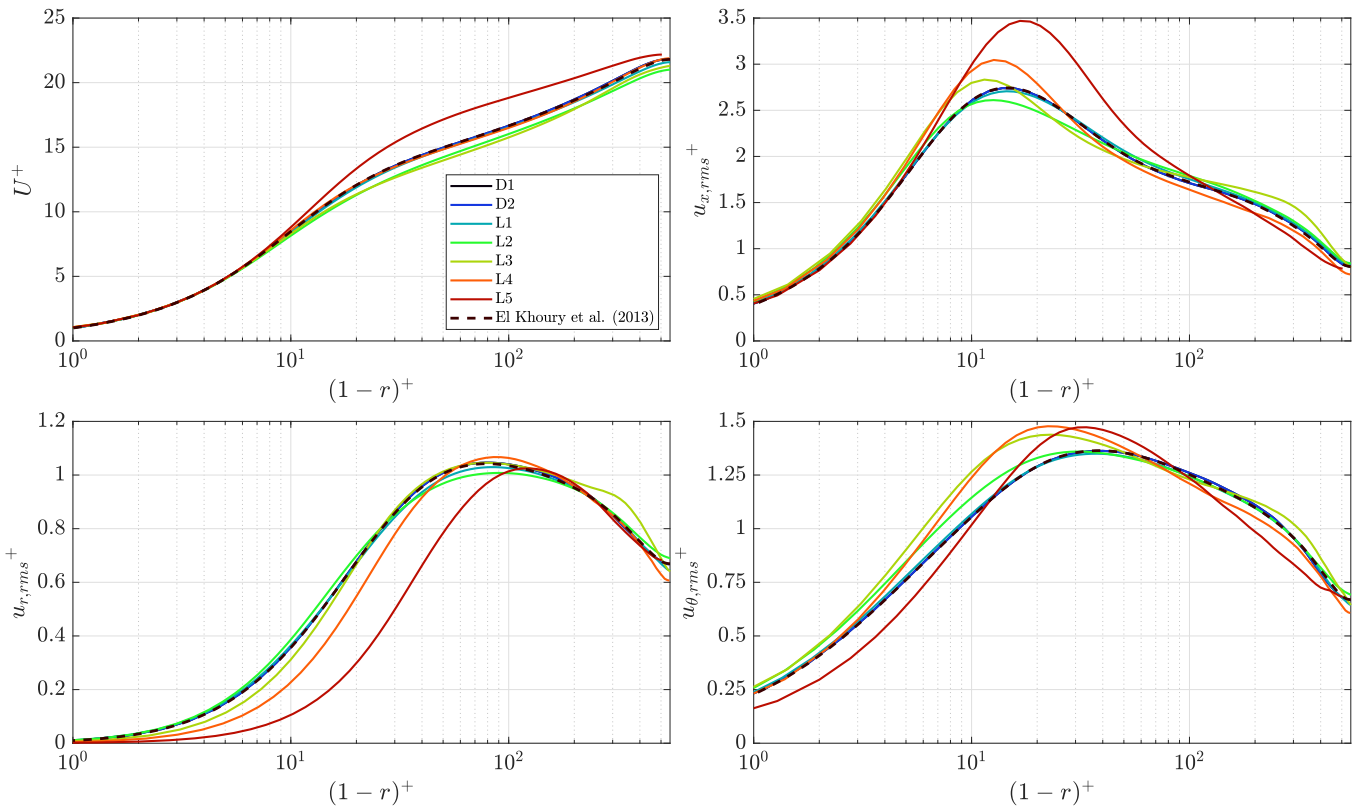}}
	\caption{Validation of the numerical simulations with reference DNS results by \citet{elkhoury2013direct}: mean flow velocity profile (top-left frame), axial (top-right frame), radial (bottom-left frame) and azimuthal (bottom-right frame) velocity profiles.}
	\label{fig:validation}
\end{figure}

The simulations were validated with reference DNS results by \citet{elkhoury2013direct}, as shown in figure \ref{fig:validation}.
Cases D1, D2 and L1 show close agreement with the reference simulations, regarding mean flow profile, axial, azimuthal and radial velocity fluctuations, almost perfectly matching the results by \citet{elkhoury2013direct}.
The coarser grid cases (L2, L3, L4 and L5) progressively deteriorate the agreement, with the L5 case showing strong mismatch with all quantities, as expected for coarser LES.

\begin{figure}
	\centerline{\includegraphics[width=\textwidth]{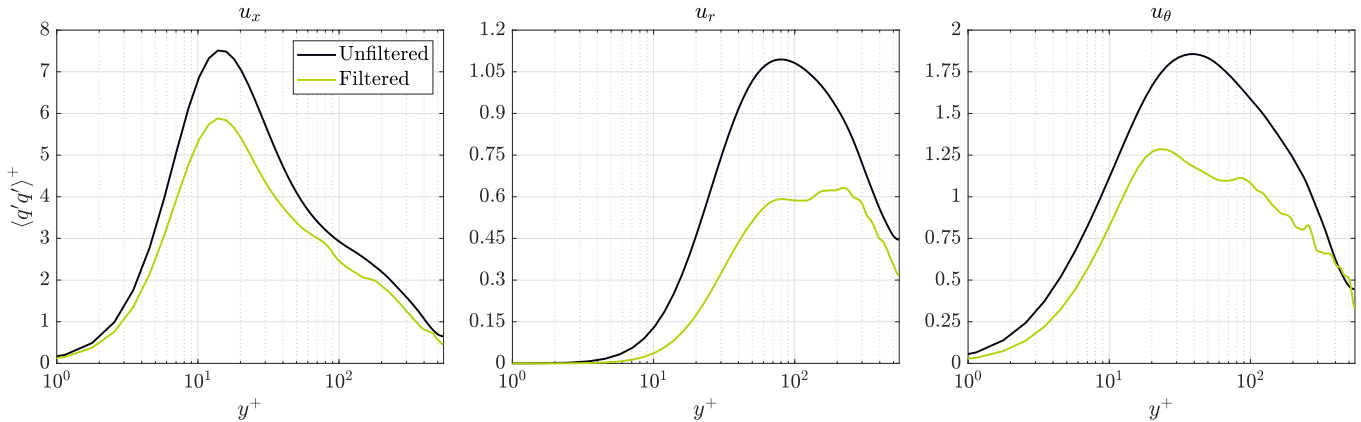}}
	\caption{Comparison between the variances of the unfiltered and filtered D2 database. Frames, from left to right: streamwise, radial and azimuthal velocity components.}
	\label{fig:variances_D2}
\end{figure}

Owing to computational cost and storage limit issues, the same downsampling strategy adopted in \citet{amaral2021resolvent} was followed here, with only the first $N_{\alpha}$ streamwise and $N_m$ azimuthal wavenumbers retained to perform the estimates, according to the values shown in table \ref{tab:simulations_cuts}.
Hence, the databases were compressed by a ratio of $N_x N_{\theta} / (N_{\alpha} N_m)$.
The filtered fields nonetheless retain the bulk of turbulent fluctuations, i.e. the peaks of the spectra in the wavenumber domain are still present after the filtering procedure.
Figure \ref{fig:variances_D2} show the variances for the unfiltered and filtered D2 database and the streamwise, radial and azimuthal velocity components, indicating that at least half of the variance of each component is retained by the filtering.
The other DNS and LES database follow similar trends.
Table \ref{tab:simulations_cuts} also displays the employed cut-off values in terms of streamwise and azimuthal wavenumbers in outer units, $\alpha_{cut}$ and $m_{cut}$, respectively, the streamwise and azimuthal wavelengths in inner/wall units, ${{\lambda_{x}}^{+}}_{cut}$ and ${{R \lambda_{\theta}}^{+}}_{cut}$, respectively, and the downsampling ratios.

\begin{table}
  \begin{center}
		\def~{\hphantom{0}}
		\caption{Wavenumber cut-off parameters used to downsample the simulations.}
		\label{tab:simulations_cuts}
		\begin{tabular}{c c c c c c c c}
				Case	& $N_{\alpha}$	&	$N_m$	& $\alpha_{cut}$	& $m_{cut}$	& ${{R \lambda_{\theta}}^{+}}_{cut}$	& ${{\lambda_{m}}^{+}}_{cut}$	& Downsampling ratio	\\ [3pt]
				\hline
				D1		& 32						& 48		& 9.425						& 47				& 365.873															& 73.368											& 182									\\
				D2		& 32						& 48		& 9.425						& 47				& 365.873															& 73.368											& 182									\\
				L1		& 32						& 48		& 9.425						& 47				& 365.873															& 73.368											& 54									\\
				L2		& 32						& 48		& 9.425						& 47				& 365.873															& 73.368											& 24									\\
				L3		& 32						& 48		& 9.425						& 47				& 365.873															& 73.368											& 14									\\
				L4		& 24						& 12		& 6.912						& 12				& 498.918															& 313.480											& 6 									\\
				L5		& 24						& 12		& 6.912						& 12				& 498.918															& 313.480											& 3										\\
		\end{tabular}
  \end{center}
\end{table}

Wall shear-stress measurements in the axial and azimuthal directions were extracted from the D1 database.
The other databases, i.e. D2, L1, L2, L3, L4 and L5, were employed to construct the estimator transfer functions by extracting the forcing statistics $\boldsymbol{P_{ff}}$ and the mean velocity profile.
We thus avoided, even for the DNS-based estimator, the use of the \emph{ground-truth} data from the D1 simulation; a separate DNS was conducted, with different initial conditions but similar statistics, to ensure that the estimator has no knowledge of the ground-truth data other than the wall measurements.
It is thus important to remark that the estimators employed in this study have no information from the D1 case.
The LES databases, which have coarser grids, significantly reduce the computational cost,  due to the lower number of grid points, but in turn lead to suboptimal estimators.
In this work we employ $\boldsymbol{P_{ff}}$ obtained in simulations that may have different grids than the measurements database D1 (see table \ref{tab:simulations_parameters}).
Hence, after the evaluation of the transfer functions $\boldsymbol{\hat{T}_q}$, they are interpolated to a grid equivalent to that of the measurements database D1.
Moreover, when the database employed to extract the forcing statistics does not contain data for a given pair $(\alpha,m)$, the forcing terms are modeled as white noise in space.
The noise CSD ($\boldsymbol{P_{nn}}$) was defined as machine precision, as we deal with DNS data mimicking measurements.

\section{Results}
\label{sec:results}

\subsection{Snapshot estimates}
\label{sec:snapshot}

\begin{figure}
	\centerline{\includegraphics[width=\textwidth]{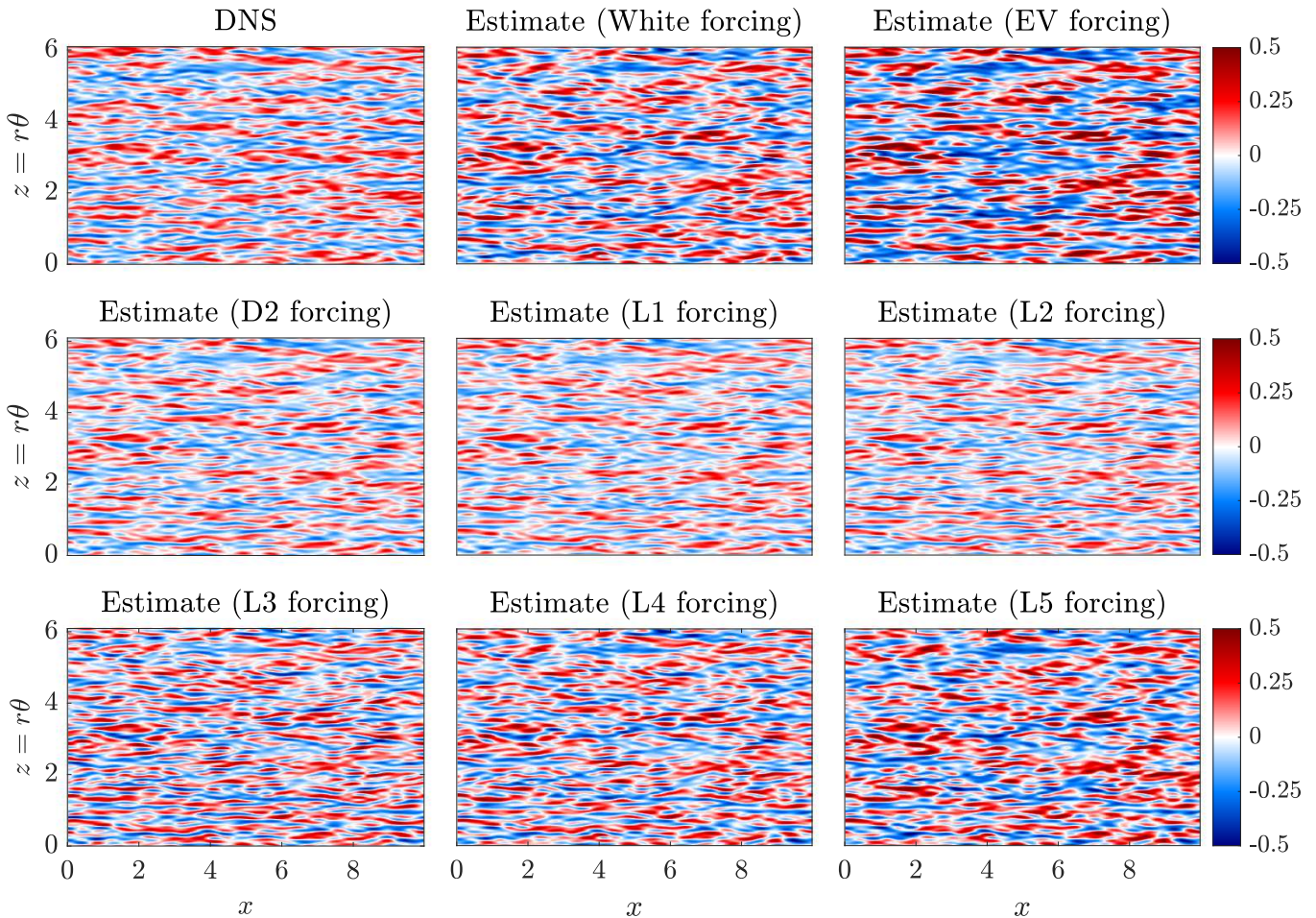}}
	\caption{Comparison between streamwise velocity component instantaneous snapshot of the D1 database and resolvent-based estimates using wall measurements of shear stress and considering white, eddy-viscosity (EV), D2, L1, L3, L4, and L5 forcing statistics at $y^+ \approx 15$. Fluctuations shown in outer units.}
	\label{fig:snapshots_yp15}
\end{figure}

\begin{figure}
	\centerline{\includegraphics[width=\textwidth]{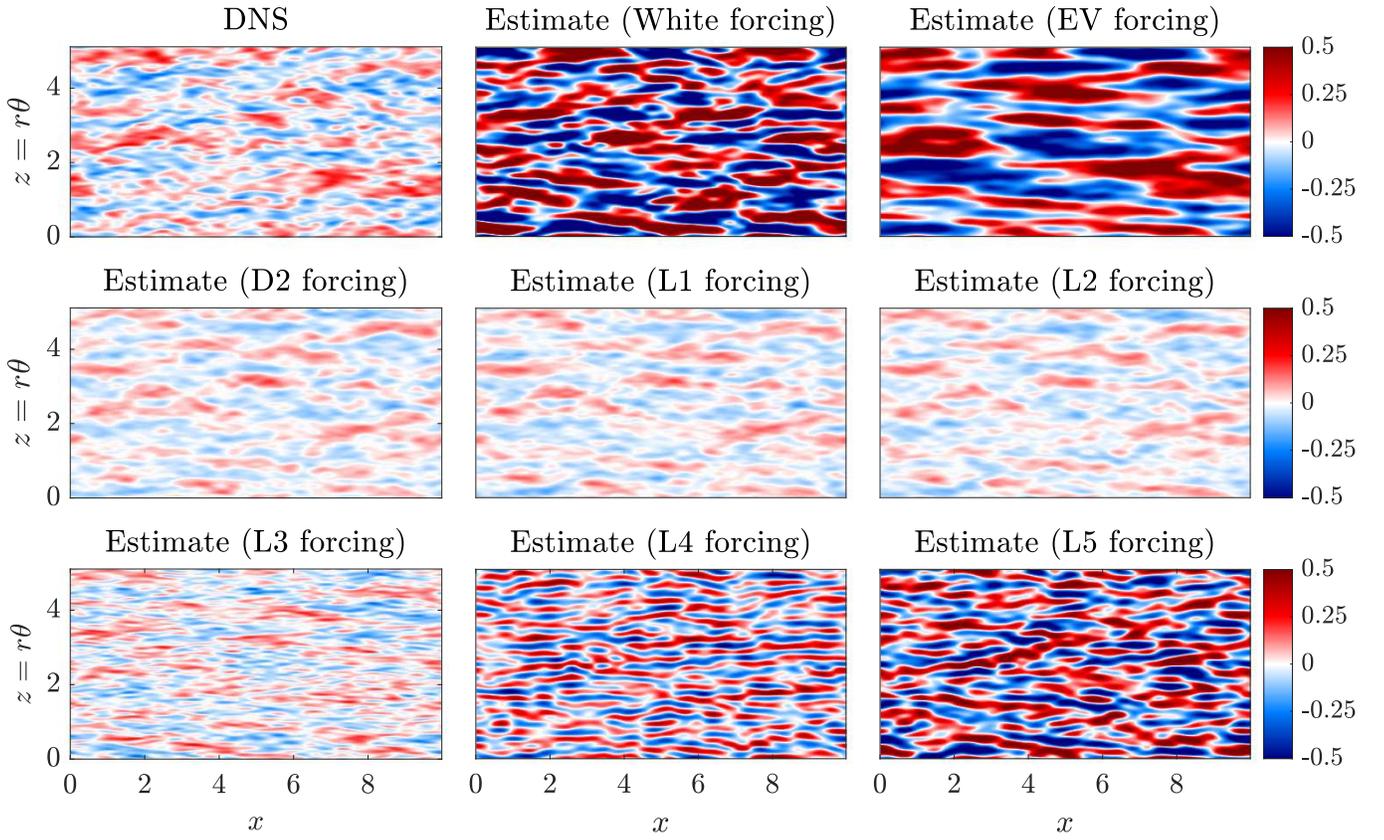}}
	\caption{Comparison between streamwise velocity component instantaneous snapshot of filtered DNS and resolvent-based estimates at $y^+ \approx 100$. See comments in the caption of figure \ref{fig:snapshots_yp15}.}
	\label{fig:snapshots_yp100}
\end{figure}

\begin{figure}
	\centerline{\includegraphics[width=\textwidth]{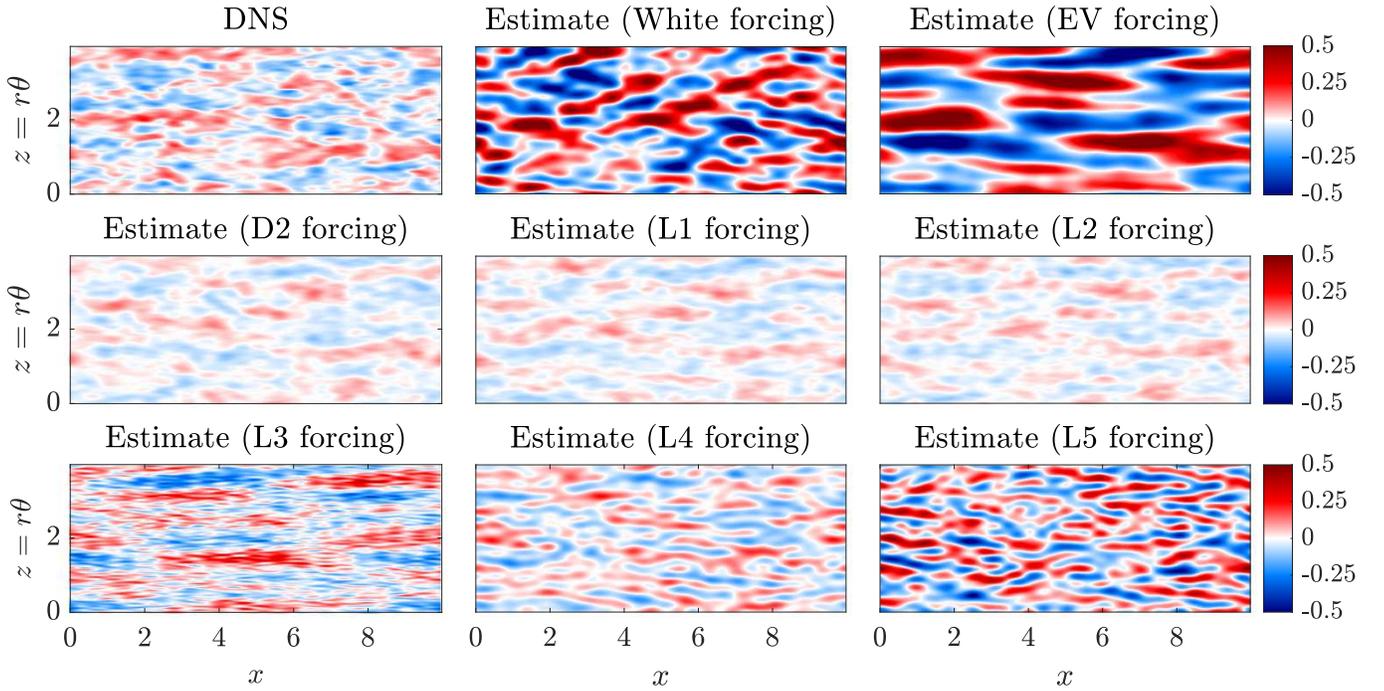}}
	\caption{Comparison between streamwise velocity component instantaneous snapshot of filtered DNS and resolvent-based estimates at $y^+ \approx 200$. See comments in the caption of figure \ref{fig:snapshots_yp15}.}
	\label{fig:snapshots_yp200}
\end{figure}

Figures \ref{fig:snapshots_yp15}, \ref{fig:snapshots_yp100} and \ref{fig:snapshots_yp200} show sample snapshots of the streamwise velocity fluctuations from the D1 database, filtered to retain only the lower axial and azimuthal wavenumbers ($N_{\alpha}$ and $N_m$), as indicated in table \ref{tab:simulations_cuts}, and corresponding estimates obtained using D2, L1, L3, L4 and L5 forcing statistics.
Results are shown at radial distances from the pipe wall of $y^+ = \left(1 - r^+\right) \approx 15$ and 100 and 200, respectively.
In the figures we use a pseudo-spanwise coordinate $z = r \theta$ ($\lambda_z = r \lambda_\theta$) for enabling comparisons with structures found in planar wall-bounded flows, in particular with the turbulent channel results of our previous work \citep{amaral2021resolvent}.
White-noise forcing statistics model (white forcing) and the linear operator containing an eddy-viscosity model (EV forcing, see appendix \ref{app:eddy_viscosity}) are also considered in the figure for means of comparison, since with these two models there is no need of prior evaluation of the forcing statistics, considered as white-noise.

When considering the buffer layer, at a wall-normal distance of $y^+ \approx 15$, the resemblance between DNS results and the estimates is remarkable, even when the white, eddy-viscosity and L5 forcing are employed.
This is somehow expected, as in the channel flow analysis in \citep{amaral2021resolvent} we have shown that assuming the forcing statistics as spatial white noise to build the estimator also provides accurate estimates for distances close to the wall.
Regarding quantitative results, as will be seen in the next section, the typical normalized error and correlation at $y^+ \approx 15$ and streamwise velocity fluctuation component are of approximately 0.5 and 0.85, respectively, for the D2, L1 and L2 estimators.
The L3, L4, L5, spatial white noise and eddy-viscosity estimators have normalized error of approximately 0.75, 0.8, 1, 0.8 and 1.05, respectively, whereas the correlations are of approximately 0.75, 0.7, 0.65, 0.7 and 0.65, respectively.
Moving further from the wall, the estimates are not as accurate, especially for the two coarser LES (L4 and L5) estimators, although most of the large-scale structures present in the DNS snapshots are still recognizable in all but the L4 and L5 estimators, as well as the white noise and eddy-viscosity estimators, in agreement with recent literature \citep{illingworth2018estimating, abreu2020resolvent, morra2021colour}.
As one moves further from the wall, only the largest turbulent structures are estimated.
For instance, as addressed in section \ref{sec:accuracy}, the normalized error and coherence at $y^+ \approx 200$ for the D1, L1 and L2 estimators and streamwise velocity fluctuation component are of approximately 0.9 and 0.4, respectively, whereas the correlations tend to zero and the errors are higher than one for coarser LES.

An interesting observation is that the well resolved LES L1 and L2 lead to estimators with similar accuracy to the optimal one, built using the DNS statistics taken from the D2 database.
This indicates that LES is a viable approach to obtain forcing statistics to build estimators of wall turbulence, without significant performance degradation if standard grid requirements for wall-resolved LES are used.
This may be understood by considering that large-eddy simulations are able to accurately resolve larger turbulent structures, which are the ones that may be estimated from the wall, as seen in earlier studies \citep{encinar2019logarithmic, wang2021state, amaral2021resolvent}.
Regarding the radial and azimuthal velocity components, not shown here, similar results as those of the streamwise velocity component were observed.

The small-scale structures can be estimated from the pipe wall even for the coarser LES, white-noise and eddy-viscosity cases, whereas large-scale structures can be observed up to $y^+ \approx 50$ only for the finer mesh cases.
Note that \citep{amaral2021resolvent} could observe structures up to $y^+ \approx 100$ using the same resolvent-based estimator and only wall-shear stress measurements, but that work built an optimal estimator using the forcing statistics extracted from the reference DNS.

\subsection{Estimate accuracy}
\label{sec:accuracy}

\begin{figure}
	\centerline{\includegraphics[width=\textwidth]{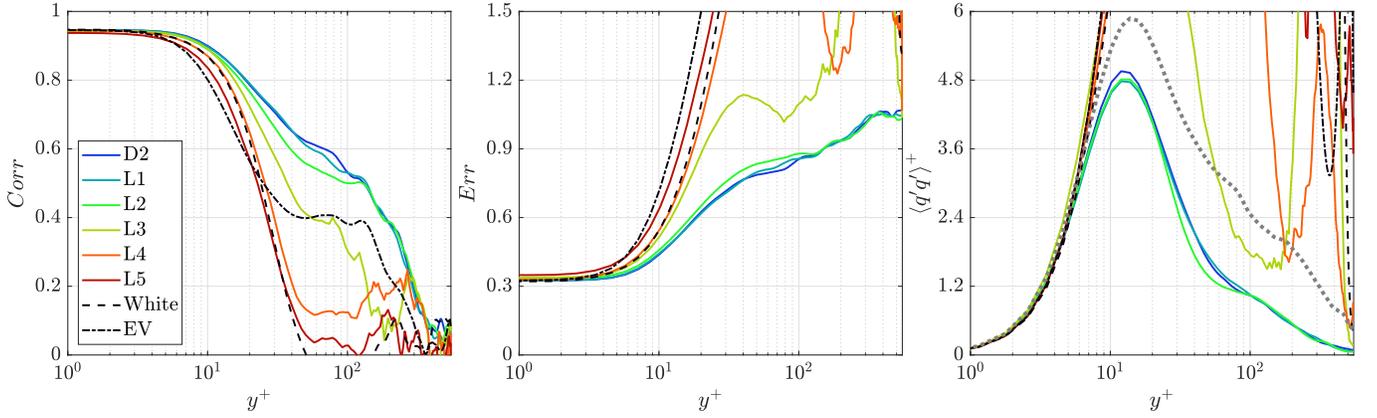}}
	\caption{Flow state comparison metrics for the streamwise velocity fluctuation component. The thick light-gray dotted line in the right frame denotes DNS (D1) variance ($\left\langle q^{\prime} {q^{\prime}} \right\rangle^{+}$) results. Frames, from left to right: correlation, normalized r.m.s. and variance. The DNS variance refers solely to wavenumbers retained for estimation.}
	\label{fig:u_metrics}
\end{figure}

Figure \ref{fig:u_metrics} displays normalized correlations ($Corr$, left frame), r.m.s errors ($Err$, middle frame) and variance ($\left\langle q^{\prime} {q^{\prime}} \right\rangle^{+}$, right frame) for the streamwise velocity component.
Correlation and error metrics are defined, as a function of wall-normal distance $y$, as
\begin{subeqnarray}
	Corr(y) &=& \frac{\int {q_{D1}}\left(y,t\right) {q_{est}}\left(y,t\right) dt}{\sqrt{\int {{q_{D1}}\left(y,t\right)}^2 dt} \sqrt{\int {{q_{est}}\left(y,t\right)}^2 dt}} \mbox{,} \\
	Err(y) &=& \frac{\sqrt{\int \left({q_{est}}\left(y,t\right) - {q_{D1}}\left(y,t\right)\right)^2 dt}}{{\sqrt{\int {q_{D1}}\left(y,t\right)^2 dt}}} \mbox{,}
	\label{eq:Corr_Err}
\end{subeqnarray}
\noindent where $q_{D1}$ denotes a flow state component, e.g. streamwise velocity fluctuation, extracted from the baseline DNS database, and $q_{est}$ denotes an estimated component, from one of the estimators considered here.

Accurate estimates correspond to low normalized error $Err$, close to 0, and high correlation $Corr$, close to 1.
A sample probe at a given $y$ position was used to evaluate the correlation and error metrics, scanning the complete time series of the baseline (D1) and estimates (D2, L1-L5, white-noise and eddy-viscosity).
The first and last $N_{fft}/2$ instants of the time series were discarded.
Such snapshots can not be estimated due to end effects when computing the convolution using finite time series.

All estimators are accurate up to $y^+ \approx 10$, showing correlation and r.m.s. errors of approximately 0.95 and 0.35, respectively, and were able to correctly identify the streaky structures of the reference DNS.
This correlation level very close to the wall is slightly lower to that obtained for $Re_{\tau} \approx 550$ turbulent channel flow estimates \citep{amaral2021resolvent}.
However, the channel estimation made use of pressure and wall shear stress from both walls, and forcing terms were directly extracted from the reference DNS, i.e. the same DNS employed for the observations of wall shear-stress and pressure.

Moving farther from the wall, only estimators D2, L1 and L2 maintain the same accuracy, especially regarding the correlation and r.m.s. metrics; correlation is higher than 0.5 up to $y^+ \approx 100$.
It is interesting that estimator L2, which has a grid with less than 10\% of the points used for the DNS-based estimator, could attain such accuracy.
This indicates that the large scales are well calculated in the LES, as expected, and their statistics may be used to build an accurate estimator at a fraction of the computational cost of the DNS-based estimator considered in \citet{amaral2021resolvent}.

\begin{figure}
	\centerline{\includegraphics[width=\textwidth]{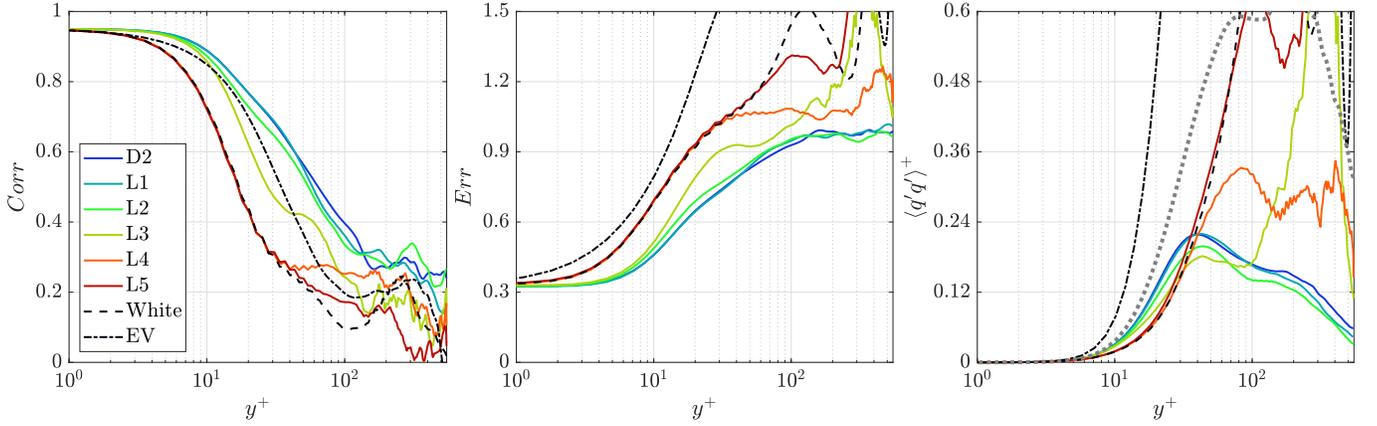}}
	\caption{Flow state comparison metrics for the radial velocity fluctuation component. See comments in the caption of figure \ref{fig:u_metrics}.}
	\label{fig:v_metrics}
\end{figure}

\begin{figure}
	\centerline{\includegraphics[width=\textwidth]{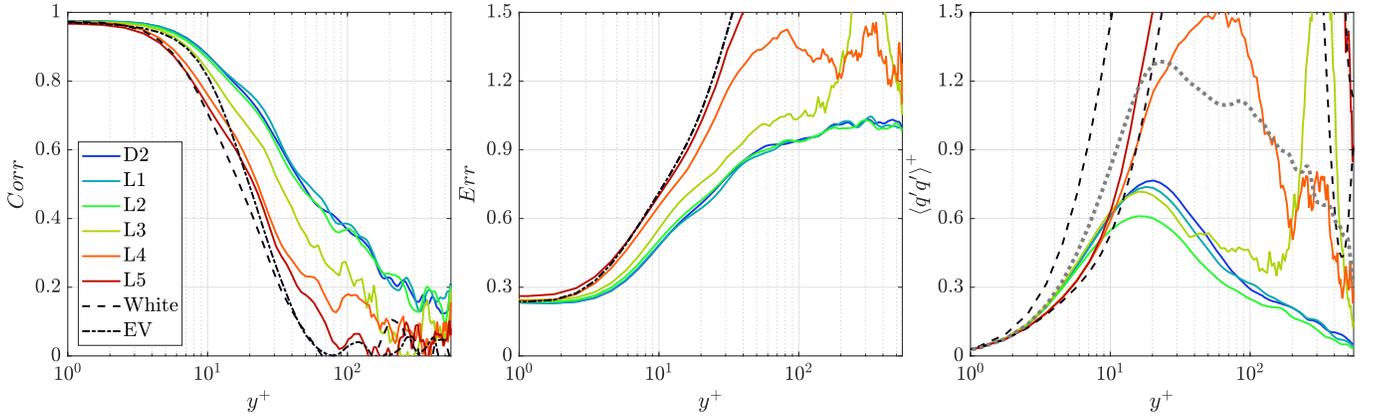}}
	\caption{Flow state comparison metrics for the azimuthal velocity fluctuation component. See comments in the caption of figure \ref{fig:u_metrics}.}
	\label{fig:w_metrics}
\end{figure}

Figures \ref{fig:v_metrics} and \ref{fig:w_metrics} show the estimates performance metrics for the radial and azimuthal velocity components, respectively.
The plots follow the same trends of the streamwise velocity component metrics, although the effects of loss of coherence, higher error and mismatch in variance for the  two coarser LES grids are more dramatic for the radial and azimuthal velocity components.
The estimators built with the finer large-eddy simulations maintain an accuracy similar to the DNS estimator based on D2.

To establish a lower-bound case of what is possible to achieve in terms of estimator performance considering cheaper forcing terms modeling, we included a case in which noise forcing statistics are modeled as spatially white (dashed lines in figures \ref{fig:u_metrics}-\ref{fig:w_metrics}) and the linear operator containing an eddy-viscosity model (dash-dotted lines in figures \ref{fig:u_metrics}-\ref{fig:w_metrics}).
It is interesting to note that in the near-wall region, up to $y^+ \approx 20$, white noise slightly outperforms the L5 estimator, and the use of L4 estimator is only justifiable above $y^+ \geq 20$.
On the other hand, the eddy-viscosity estimator outperforms even the L4 estimator for $y^+ \geq 55$, corroborating previous studies \citep{amaral2021resolvent} that showed the eddy-viscosity model is a good option to model the nonlinear terms of the Navier-Stokes system for distances far from the wall.
Overall, the quantitative metrics in figures \ref{fig:u_metrics}-\ref{fig:w_metrics} confirm the qualitative results shown in figures \ref{fig:snapshots_yp15}-\ref{fig:snapshots_yp200}, although for the azimuthal and radial velocity components the eddy-viscosity model can even outperform the L4 estimator, but not the better resolved LES estimators.

\begin{figure}
	\begin{subfigure}{\textwidth}
		\centerline{\includegraphics[width=\textwidth]{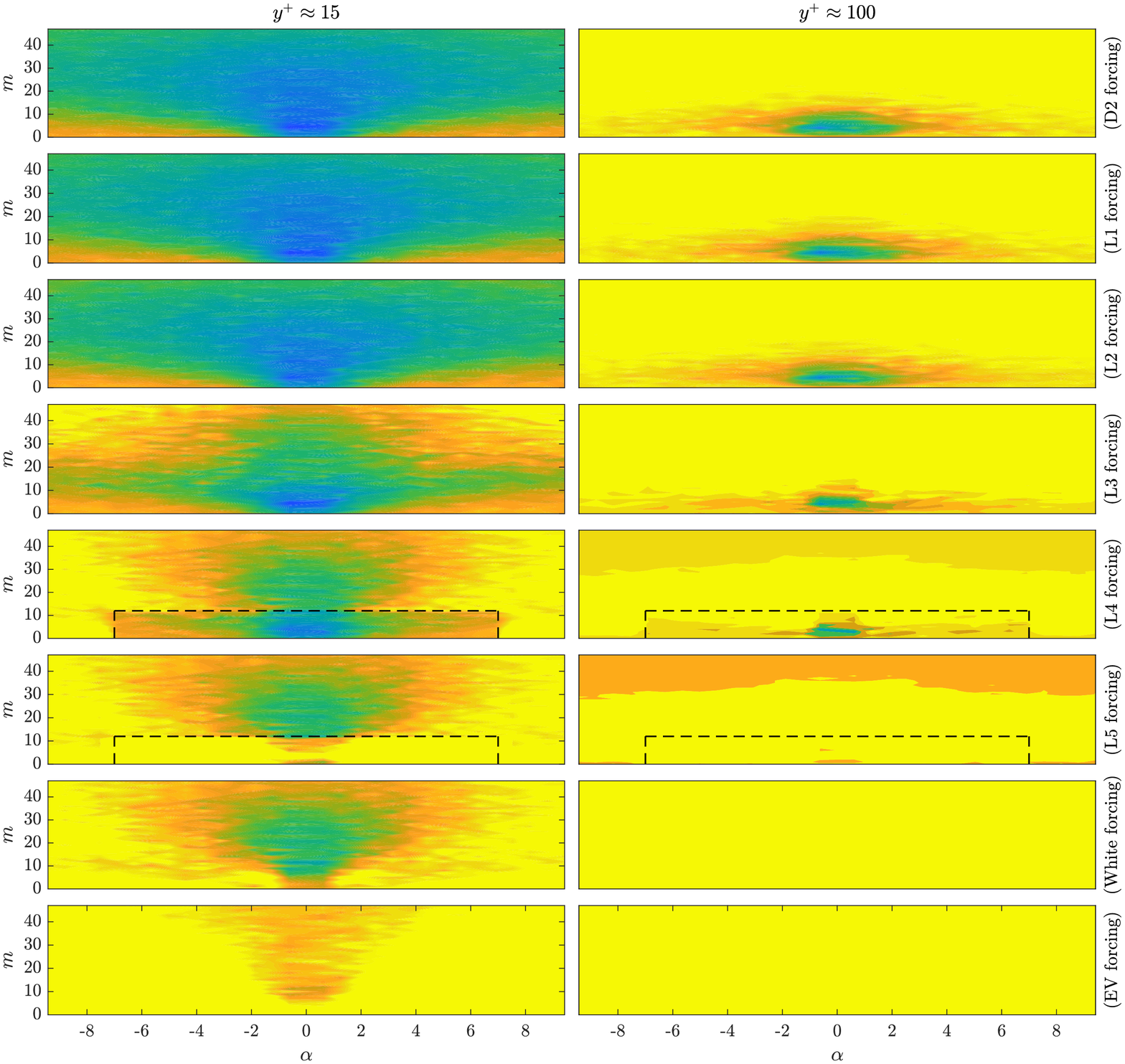}}
	\end{subfigure}
	\begin{subfigure}{\textwidth}
		\vspace{2.5mm}
		\centerline{\includegraphics[width=0.8\textwidth]{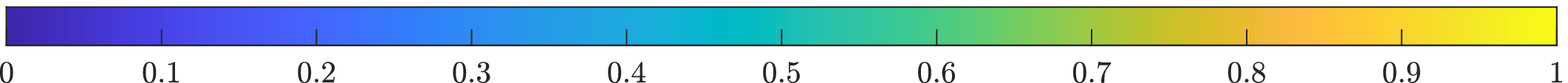}}
	\end{subfigure}
	\caption{Normalized r.m.s. error as a function of $(\alpha,m)$ state comparison metrics for D2, L1, L2, L3, L4, L5, white-noise and eddy-viscosity estimators. Normalized errors higher than 1 are saturated.}
	\label{fig:alpha_beta_error}
\end{figure}

Normalized r.m.s. error as a function of the wavenumber pair $(\alpha,m)$ and wall-distance $y$ are shown in figure \ref{fig:alpha_beta_error} for wall distances of $y^+ \approx 15$ and 100.
The error metric is defined as
\begin{equation}
	Err(\alpha,y,m) = \frac{\sqrt{\int \sum_{i=1}^{3} \left|{q_{est}^i}\left(\alpha,y,m,t\right) - {q_{D1}^i}\left(\alpha,y,m,t\right)\right|^2 dt}}{{\sqrt{\int \sum_{i=1}^{3} \left|{q_{D1}^i}\left(\alpha,y,m,t\right)\right|^2 dt}}} \mbox{,}
	\label{eq:RMSerror_alpha_beta}
\end{equation}
\noindent with superscript $i$ in ${q^{i}}$, for $i = 1$, 2 and 3, denoting streamwise (${u_x}^{\prime}$), wall-normal/radial (${u_r}^{\prime}$) and azimuthal (${u_\theta}^{\prime}$) velocity fluctuation components, respectively.
In this figure, the normalized r.m.s errors are shown in the 0 to 1 range and errors higher than 1 are saturated.

Note that for cases L4 and L5 (rows 5 and 6 of figure \ref{fig:alpha_beta_error}) the region delimited by the dashed lines is the region on which the streamwise and azimuthal wavenumbers are contained within the LES databases.
For the streamwise and azimuthal wavenumbers outside that region, the white noise estimator is considered in the calculations, as can be observed on row 8 of figure \ref{fig:alpha_beta_error}.

The large structures, which are characterized by small $\alpha$ and $m$, are accurately estimated for the D2, L1 and L2 cases, with virtually zero r.m.s. error at both planes.
The estimates for smaller structures (large $\alpha$ and $m$), on the other hand, display higher r.m.s. error, especially for the $y^+ \approx 100$ plane.
For the coarser L4 and L5 estimators, even for the $y^+ \approx 15$ plane, the accuracy of smaller structure estimates is quite low.
This indicates that such very coarse LES are not suitable to build estimators; notice that the grid spacing of L5, of about 100 wall units in the azimuthal direction, is close to the typical streak spacing; hence, near wall structures cannot be captured by the coarser LES, which compromise their potential to build resolvent-based estimators.
However, the finer LES allow an accuracy close to the one from the optimal D2 estimator, showing that LES is a viable approach to obtain forcing statistics required for resolvent-based estimation.
This is far superior than the estimates built from white-noise assumption, also shown in figure \ref{fig:alpha_beta_error}, which have a large error as one moves away from the wall. 
Inclusion of an eddy viscosity in the operator, also shown in figure \ref{fig:alpha_beta_error}, and consideration of white-noise forcing also leads to inaccurate estimates.
This finding is in agreement with recent literature \citep{symon2021energy} that showed the use of eddy-viscosity models may lead to errors regarding the modeling of turbulent structures.

Regarding the LES requirements to obtain reasonable accurate results, it is observed that simulations with mesh discretization of approximately $\Delta x^+ \leq 20$ and $\Delta z^+ \leq 12$ provided the best estimates, which agrees with the standards for wall-resolved simulations \citep{rasam2011effects, encinar2019logarithmic}.
In other words, we verify here that it is necessary to have a well-resolved LES near the walls to obtain good estimates from measurements obtained at the walls.
Our estimates up to the L2 case have error magnitudes that are overall lower than the results presented by \citet{illingworth2018estimating} and \citet{oehler2018linear}, which employed an eddy-viscosity model and a Kalman filter to estimate turbulent channel flows at $Re_\tau = 1000$ and 2000, respectively, from streamwise and spanwise velocity measurements; this highlights that the use of forcing statistics from wall-resolved LES is a viable approach to build more accurate estimators.
The present results are in agreement with our earlier channel results \citep{amaral2021resolvent}, with the caveat that we are dealing with turbulent pipe flow instead of the channels studied in the cited works.

\subsection{Structures observable from the pipe wall}
\label{sec:footprint}

\citet{smits2011high} and \citet{jimenez2013near}, among other authors, have shown that, for wall-bounded flows, many structures leave their footprint on the walls, even the ones present in the outer layer.
In order to explore which flow structures leave a footprint on the walls, in \citet{amaral2021resolvent} we introduced a metric that consists of a distance from the wall, the maximum observed height, $y_{obs} = (1 - r_{obs})$, from which the estimate normalized error is $Err(\alpha,y_{obs},m) \leq 0.5$.
We select an error value of 0.5, considering that an estimator that attains 50\% of accuracy or more is good enough, but this value can be calibrated with a more restrictive criterion if desired.
As observed in the previous results, the estimators used in this study based on wall-measurements of shear stress lose accuracy far from the wall, and only the largest scales may be estimated from wall measurements.

\begin{figure}
	\centerline{\includegraphics[width=\textwidth]{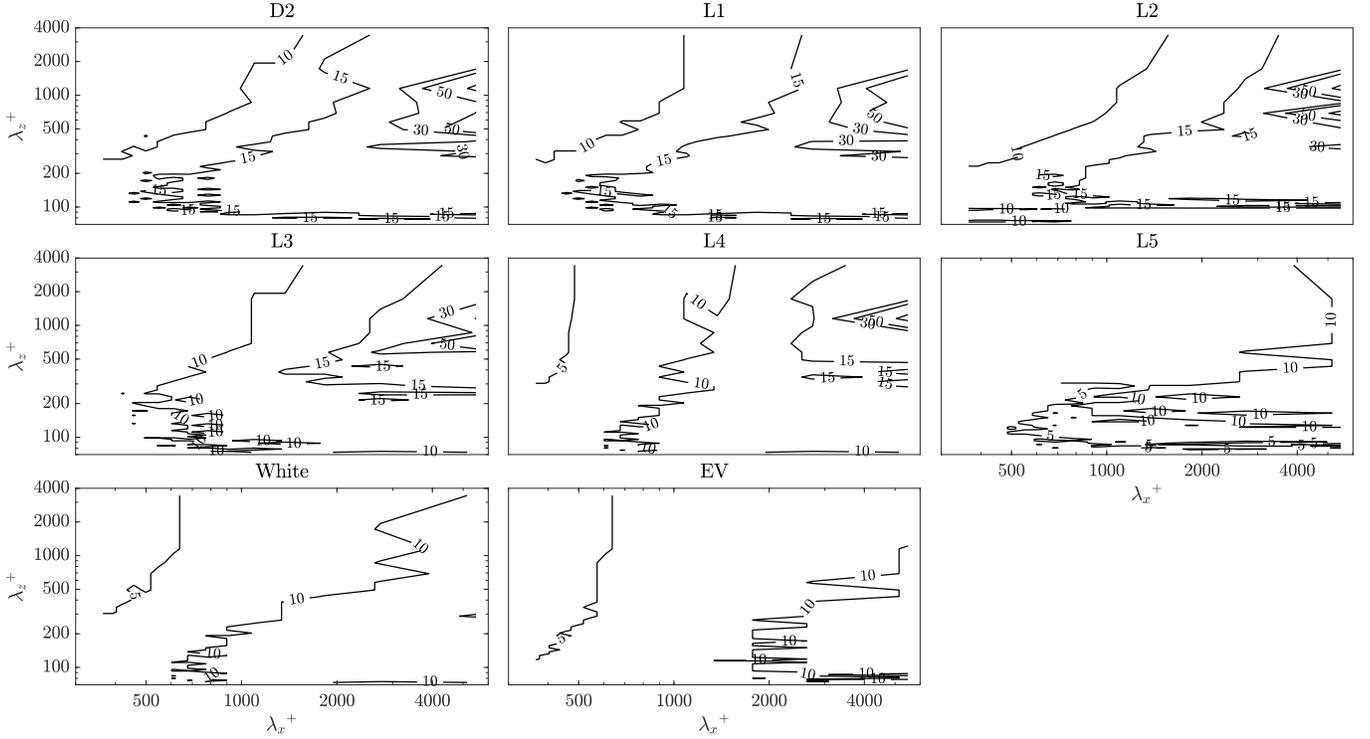}}
	\caption{Observed height in wall/inner units (${y_{obs}}^{+} = Re_{\tau} - {r_{obs}}^{+}$) for D2, L1, L2, L3, L4, L5, white-noise and eddy-viscosity estimators.}
	\label{fig:yobs}
\end{figure}

Figure \ref{fig:yobs} shows the maximum observed height in plus units (${y_{obs}}^{+}$) as a function of streamwise and pseudo-spanwise wavelengths (${\lambda_x}^+$,${\lambda_z}^+$), with $\lambda_z = r \lambda_\theta$.
Contour levels of $y^{+} = 5, 10, 15, 30$ and 50 are displayed in the maps.
For all estimators, the smaller structures (of small wavelength pairs) can only be well estimated very close to the wall, whereas far from the wall only large structures (of large wavelength pairs) can attain some level of accuracy.
Estimators D2, L1, L2 and L3 display similar behavior, keeping accuracy for large structures up to ${y}^{+} \approx 50$.

The worse performance of estimators L4 and L5 is mostly associated to smaller structures, which cannot be accurately resolved by the coarser grids.
It is interesting to notice that for larger structures, $({\lambda_x}^+,{\lambda_z}^+) \approx (4000,~1000)$, even the coarse-LES estimator L4 maintains some accuracy of estimates, which is in line with the idea that even coarse LES are able to resolve the largest turbulent structures \citep{rasam2017improving}.
However, if the grid is too coarse there is a worsening of estimates at all scales, as observed from the L5 results, which show performance lower than the white-noise and eddy-viscosity estimators.

\section{Conclusions}
\label{sec:conclusions}

In this paper we employed a (linear) resolvent-based estimator methodology \citep{martini2020resolvent} to obtain the space--time flow field of a $Re_{\tau} \approx 550$ turbulent pipe flow from wall-shear stress low-rank measurements.
A DNS database was used to extract the wall-shear stress measurements, whereas other DNS and LES databases were employed for the modeling of the statistics of non-linear terms (which constitute a forcing in resolvent analysis) necessary for state estimation.
Hence, an optimal estimator, built from DNS statistics, is compared to sub-optimal resolvent-based estimators informed by LES.

We compared the accuracy of the estimators in terms of snapshot reconstruction, correlation, normalized error and variance.
Satisfactory results were obtained with the forcing statistics from LES, especially up to the buffer layer.
The accuracy progressively deteriorates for distances far from the wall and for coarser LES meshes, although for cases D2, L1 and L2 the large-scale structures can still be recognizable up to $y^+ \approx 200$.
The LES-based estimators using typical grids for wall turbulence maintain an accuracy similar to the optimal estimation built from DNS statistics (D2).
The present results are in agreement with recent literature on estimation of wall-bounded flows from wall measurements.
For instance, \citet{encinar2019logarithmic}, estimated turbulent structures of channel flows for $932 \leq Re_\tau \leq 5300$ regimes using a linear stochastic estimator.
The authors attained a good level of accuracy near the walls, i.e. at $y/H \approx 0.2$, where $H$ is the channel height.
\citet{guastoni2021convolutional}, on the other hand, employed two different convolutional-network algorithms to estimate the instantaneous velocity components of channel flows at $Re_\tau = 180$ and 550, obtaining good agreement with reference result up to $y^+ = 50$.
The question that remains is that at what computational cost the two strategies above cited can attain the same accuracy as the resolvent-based estimator employed together with forcing statistics provided by LES.

The accuracy level of a LES estimator containing approximately 10\% of the grid points of the DNS database, case L3 (see tables \ref{tab:simulations_parameters} and \ref{tab:simulations_cuts}), which is very close to what is obtained with the DNS estimator, case D2.
It would be desirable to be able to estimate flow fluctuations from wall measurements with a simple model for forcing statistics, but the consideration of white-noise forcing leads to inaccurate estimates; hence, \emph{linear} estimation requires information on the statistics of \emph{non-linear} terms.
The present results show that such statistics of non-linearity do not need unrealistic levels of accuracy, and moderately coarse large-eddy simulation may provide the required information on dominant non-linear effects.
LES-informed resolvent-based estimation is thus a viable approach for accurate estimates of turbulent flow at high Reynolds numbers and can also be seen as a more accurate alternative, with moderate additional cost, to the use of eddy-viscosity models that have been recently explored in the literature for wall-bounded flows \citep{delalamo2006linear, morra2019relevance, hwang2010linear, pickering2021optimal}.




\noindent{\bf Funding\bf{.}} F. R. Amaral received funding from from São Paulo Research Foundation (FAPESP/Brazil), grant \#2019/02203-2. A. V. G. Cavalieri was supported by the National Council for Scientific and Technological Development (CNPq/Brazil), grant \#313225/2020-6. The authors were also funded by FAPESP/Brazil, grant \#2019/27655-3.\\

\noindent{\bf  Author ORCID\bf{.}} F. R. Amaral, https://orcid.org/0000-0003-1158-3216; A. V. G. Cavalieri, https://orcid.org/0000-0003-4283-0232\\


\noindent{\bf Declaration of Interests\bf{.}} The authors report no conflict of interest. \\


\appendix

\section{Eddy viscosity model}
\label{app:eddy_viscosity}

If an eddy-viscosity model is considered, the LNS equations can be written as
\begin{subeqnarray}
	{\partial_t \boldsymbol{u}} + {u_r \partial_r \bar{U} \boldsymbol{e_x}} + {\bar{U} \partial_x \boldsymbol{u}} &=& {\boldsymbol{\nabla} p} + \frac{1}{Re} {\boldsymbol{\nabla} \cdot \left[\nu_T\left(r\right) \left(\nabla \boldsymbol{u} + \nabla \boldsymbol{u}^{T}\right)\right]}) + \boldsymbol{f} \mbox{,} \\
	\boldsymbol{\nabla} \cdot \boldsymbol{u} &=& 0 \mbox{.}
	\label{eq:LNS_eddy_viscosity}
\end{subeqnarray}
\noindent where $\nu_T$ is the eddy-viscosity, which can be modeled as \citep{cess1958survey}
\begin{equation}
	\frac{\nu_T(y)}{\nu} = \frac{1}{2} \left\{1 + \frac{\kappa^2 {\hat{Re}}^2 \hat{B}}{9} \left(2 y - y^2\right)^2 \left(3 - 4 y + 2 y^2\right)^2 \left[1 - e^{\left(\frac{- y \hat{Re} \sqrt{\hat{B}}}{A^{+}}\right)}\right]^2\right\}^{1/2} + \frac{1}{2} \mbox{,}
	\label{eq:eddy_viscosity_model}
\end{equation}
\noindent where $y = 1 - r$,the constants $\kappa$ and $A^{+}$ are given as 0.42 and 27, respectively \citep{mckeon2005new}, $\hat{Re} = Re/2$ and $\hat{B} = - 2 \partial_x p$.

Following \citet{willis2010optimally}, the linear operator accounting for the eddy-viscosity model is given by
\begin{equation}
	\boldsymbol{A} =
	\left[\begin{array}{cccc}
		- i \alpha \boldsymbol{\bar{U}} + \frac{1}{Re}{\left(\nu_T\left(\boldsymbol{\Delta} + \boldsymbol{r}^{-2}\right) + \boldsymbol{E} \right)}	& - \boldsymbol{D}\boldsymbol{\bar{U}} + \frac{1}{Re}{i \alpha {\nu_T}^{\prime}}	& \boldsymbol{Z}	& - i \alpha \boldsymbol{I} \\
		\boldsymbol{Z}	& - i \alpha \boldsymbol{\bar{U}} + \frac{1}{Re}{\left({\nu_T \boldsymbol{\Delta}} + {2 \boldsymbol{E}}\right)}	& - {\frac{1}{Re} \boldsymbol{F}}	& - \boldsymbol{D} \\
		\boldsymbol{Z}	& \frac{1}{Re}{\left(\boldsymbol{F} + {i \beta {\nu_T}^{\prime} \boldsymbol{r}^{-1}}\right)}	& - i \alpha \boldsymbol{\bar{U}} + \frac{1}{Re}\left({\nu_T \boldsymbol{\Delta}} + \boldsymbol{G}\right) & - i \beta \boldsymbol{r}^{-1}	\\
		i \alpha \boldsymbol{I}	& \boldsymbol{D} + \boldsymbol{r}^{-1}	& i \beta \boldsymbol{r}^{-1}	& \boldsymbol{Z}
	\end{array}\right]
	\label{eq:linear_operator_eddy_viscosity} \mbox{.}
\end{equation}
\noindent where
\begin{subeqnarray}
	\boldsymbol{\Delta} = \boldsymbol{\nabla}^2 - \boldsymbol{r}^{-2} &=& - {\alpha^2 \boldsymbol{I}} - {\left(\beta^2 + 1\right) \boldsymbol{r}^{-2}} + {\boldsymbol{r}^{-1} \boldsymbol{D}} + \boldsymbol{D}^2 \mbox{,}	\\
	{\nu_T}^{\prime} &=& \boldsymbol{D} {\nu_T} \mbox{,}	\\
	\boldsymbol{E} &=& {\nu_T}^{\prime} \boldsymbol{D} \mbox{,}	\\
	\boldsymbol{F} &=& 2 i \beta \nu_T \boldsymbol{r}^{-2} \mbox{,}	\\
	\boldsymbol{G} &=& {\nu_T}^{\prime} \left(\boldsymbol{D} - \boldsymbol{r}^{-1}\right) \mbox{.}
	\label{eq:viscous_operators}
\end{subeqnarray}

\section{Block size convergence tests}
\label{app:convergence}

\begin{figure}
	\centerline{\includegraphics[width=\textwidth]{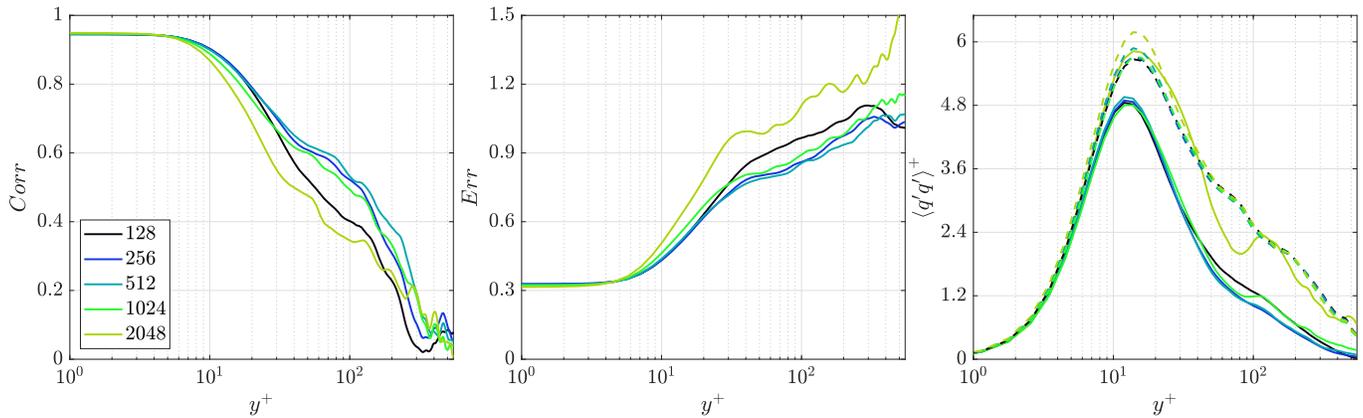}}
	\caption{Flow state comparison metrics for the streamwise velocity fluctuation component and different block sizes using D2 forcing statistics. Dashed curves in the left frame denote DNS (D1) results. Frames, from left to right: correlation, normalized r.m.s. and variance. The DNS variance refers solely to wavenumbers retained for estimation.}
	\label{fig:nfft}
\end{figure}

Figure \ref{fig:nfft} displays a convergence test for various block sizes ($128 \leq N_{fft} \leq 2048$) while keeping the block overlap fixed in 75\%.
The metrics depicted in the figure are the correlation, the normalized error and the variance of the streamwise velocity fluctuation component.
To obtain these results, the forcing terms were extracted from case D2.
It is observed that the statistics are well converged in the $256 \leq N_{fft} \leq 1024$ range.
This shows the typical compromise in the application of the Welch method to obtain frequency-domain statistics: blocks that are too short have a low frequency resolution, while long blocks may lead to worse statistical convergence.
It is nonetheless reassuring that various signal processing choices lead to similar estimation properties.
We have taken the intermediate value $N_{fft} = 512$ to obtain the forcing statistics used to build the estimators.
Similar results, not shown here, were obtained for the radial and azimuthal velocity fluctuation components.

\FloatBarrier


\bibliographystyle{unsrtnatabbrv}
\bibliography{references}

\end{document}